\documentclass[12pt, letterpaper]{article}

\pdfoutput=1

\usepackage{arxiv}
\usepackage[utf8]{inputenc} 
\usepackage[T1]{fontenc}    
\usepackage{amsmath,amsfonts,amssymb,bm}
\usepackage{mathtools}
\usepackage{mathptmx}
\usepackage{newtxtext}
\usepackage{newtxmath}
\usepackage{hyperref}
\usepackage{caption}
\usepackage{float}
\usepackage{xcolor}
\usepackage{multirow}
\usepackage{subcaption,booktabs}
\usepackage{rotating}
\usepackage{adjustbox}
\usepackage{graphicx}
\usepackage{array}
\usepackage{environ}
\usepackage[square,sort,comma,numbers]{natbib}

\title{Good Practices and Common Pitfalls in Climate Time Series Changepoint Techniques: A Review}

\author{
Robert Lund \\
Department of Statistics,\\
University of California, Santa Cruz \\
  \texttt{rolund@ucsc.edu} \\
 \And
Claudie Beaulieu \\
Department of Ocean Sciences, \\
University of California, Santa Cruz\\
  \And
Rebecca Killick \\
Department of Statistics,\\
 Lancaster University, UK\\
  \And
QiQi Lu \\
Department of Statistical Sciences and Operations Research, \\
Virginia Commonwealth University\\
  \And
Xueheng Shi \\
Department of Statistics,\\
University of California, Davis \\
\texttt{xhshi@ucdavis.edu}
}

\begin{document}

\maketitle

\begin{abstract}
Climate changepoint (homogenization) methods abound today, with a myriad of techniques existing in both the climate and statistics literature.  Unfortunately, the appropriate changepoint technique to use remains unclear to many. Further complicating issues, changepoint conclusions are not robust to small perturbations in assumptions; for example, allowing for a trend or correlation in the series can drastically change conclusions. This paper is a review of the changepoint topic, with an emphasis on illuminating the models and techniques that allow the scientist to make reliable  conclusions.  Pitfalls to avoid are demonstrated via actual applications.  The discourse begins by narrating the salient statistical features of most climate time series.  Thereafter, single and multiple changepoint problems are considered.  Several pitfalls are discussed en route and good practices are recommended.  While the majority of our applications involve temperature series, other settings are mentioned.
\end{abstract}

\section{Introduction}

Climate time series often contain nonlinearities or sudden structural changes in their behavior. Such features may reflect linear or nonlinear dynamics in the climate system, and need to be detected and modeled for an accurate depiction of long-term changes in the time series \citep{Beaulieu_2012, Beaulieu_Killick_2018, Cahill_etal_2015, Mudelsee_2019}. Alternatively, these features may reflect artificial discontinuities induced by changes in measurement practices (e.g., station relocations, gauge changes, observer changes) \citep{Menne_Williams_2009, Ribeiro_2016, Peterson_al_1998, Venema_etal_2012}. Some (but not necessarily all) of these artificial changes induce shift discontinuities into the series. If these shifts are not detected and removed from the data, results of analysis using these data may be biased or erroneous. Regardless of the underlying cause of the shift, if the true number of changes and their timings are unknown, changepoint detection techniques are typically used to estimate these key quantities. If the change is artificial, the number of changepoints and their locations are needed to adjust (homogenize) climate records {\em a priori} for realism. If the change is caused by variability or forcings in the climate system,  the number of changepoints and their timings are needed to accurately represent long-term changes in the data.

Changepoint detection is a rapidly growing field in the statistical literature, and applications to climate time series are numerous. This paper contains a modern statistical review of the changepoint topic in climate settings. Our overarching goal is to develop a good estimate of the number of changepoints and their locations, and to accessibly present the methods for the climate scientists and experts with a minimum of jargon and technicalities (some technical methods, of course, are needed).  The paper intends to serve as a technical guide about changepoint detection, informing the researcher of the appropriate methods to use given statistical properties of the time series. Unfortunately, changepoints are a thorny statistical issue: small changes in assumptions can yield very different fitted changepoint configurations (e.g., \cite{Lund_Reeves_2002, Beaulieu_2012, Beaulieu_Killick_2018}).  Because of this, it is important for researchers to be aware of changepoint/homogenization pitfalls.  This paper aims to illuminate some common changepoint mistakes, and to make recommendations on the best practices to avoid them.

Even in a review paper such as this, concessions must be made for length.  In particular, this paper will not compare or classify the many software packages used today to homogenize climate time series. Indeed, our focus is on the techniques themselves, the intent being to illuminate the concepts that underlie sound changepoint analyses.  The paper will not delve into attribution of any discovered changepoints in our examples --- what caused the changepoints is immaterial in this discussion.  Toward this, most homogenizations only want to remove changepoint features from the record that can be attributed to man-made (artificial) influences --- changepoints caused by natural fluctuations should be retained. This is best done by subtracting references series from nearby locations from the target series to be homogenized before analysis to eliminate naturally occurring fluctuations.  These so-called absolute versus relative homogenization procedures, and the ``target'' and ``reference'' series involved in them, are discussed in \cite{Menne_Williams}.  While a bit more is said about this in the next section, the issue is not a prominent feature of this paper. 

The rest of this paper proceeds as follows.  The next section discusses  the statistical properties of typical climate time series, delving into correlation, trends, seasonality, and changepoints.  Here, target and reference series are introduced and absolute versus relative homogenization procedures are distinguished.  Section 3 introduces a time series regression model that describes a wide suite of climate series.  This model provides the mathematical backdrop for all changepoint analyses.  Section 4 considers the case of a single changepoint, presenting what is generally viewed as the best (most powerful) single changepoint detector.  Section 5 moves to the multiple changepoint case, which arises when one does not {\em a priori} know how many changepoints are present, the typical setting in practice.  Section 6 then transitions to a list of pitfalls to avoid when homogenizing climate series. Section 7 closes with conclusions and comments, including avenues for future research.

\section{Statistical Properties of Climate Time Series}

\begin{figure}[ht]
	\centering
	\noindent\includegraphics[width=\textwidth,angle=0]{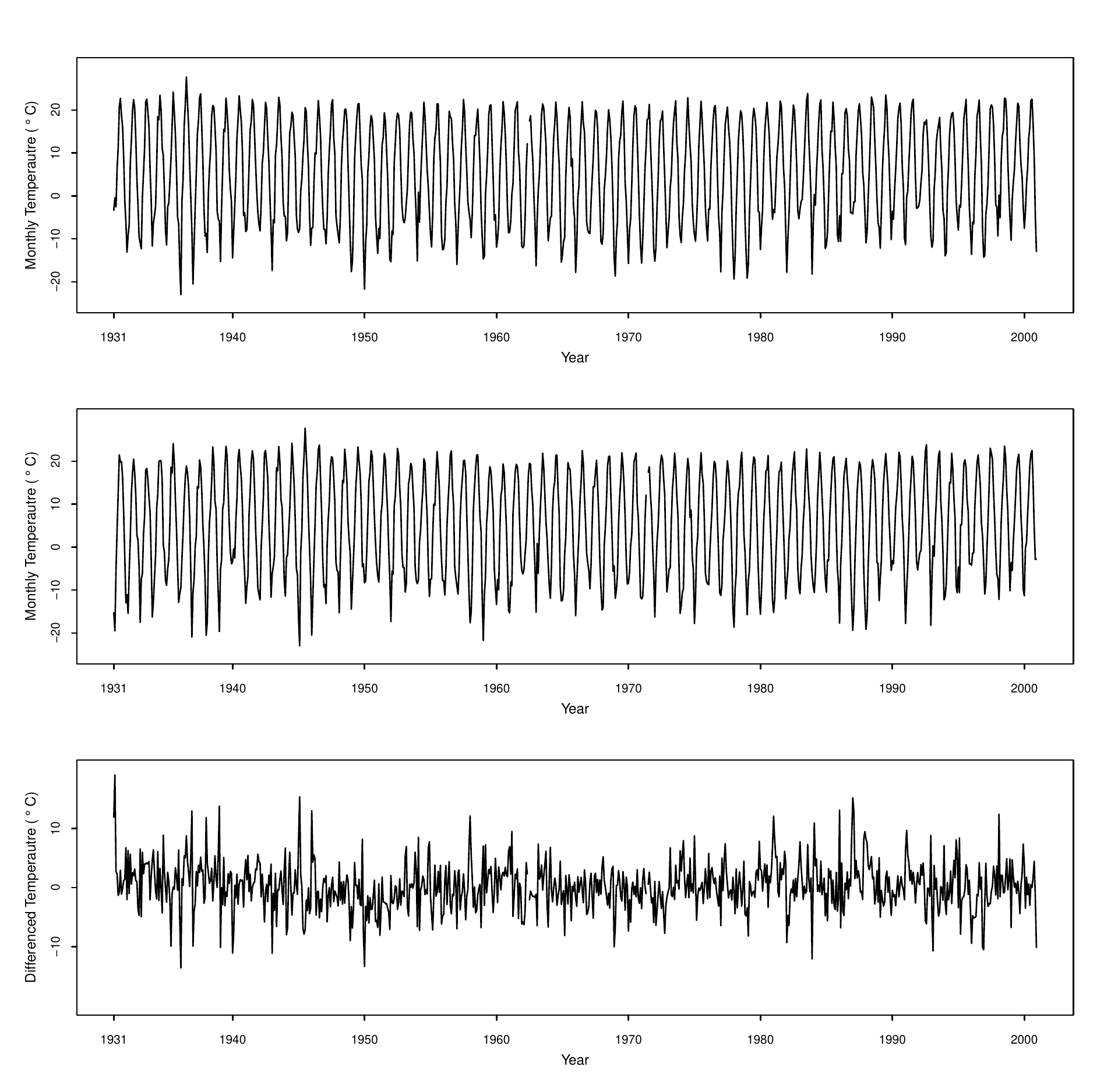}\\
	\caption{Monthly averaged temperatures at the Mott (Top) and 
		Richardton-Abbey (Middle) stations in west-central North Dakota.  The bottom graphic shows the Mott minus Richardton-Abby series in a target minus reference comparison.}
	\label{NDdata}
\end{figure}

Figure \ref{NDdata} presents 71 years of monthly averaged temperatures from two nearby stations in west central North Dakota: Mott and Richardton-Abby. The records span January, 1931 --- December, 2001.  These series will serve to illustrate our list of salient statistical features in climate series.

\subsection{Seasonality}

\begin{figure}[ht]
	\centering
	\noindent\includegraphics[width=\textwidth,angle=0]{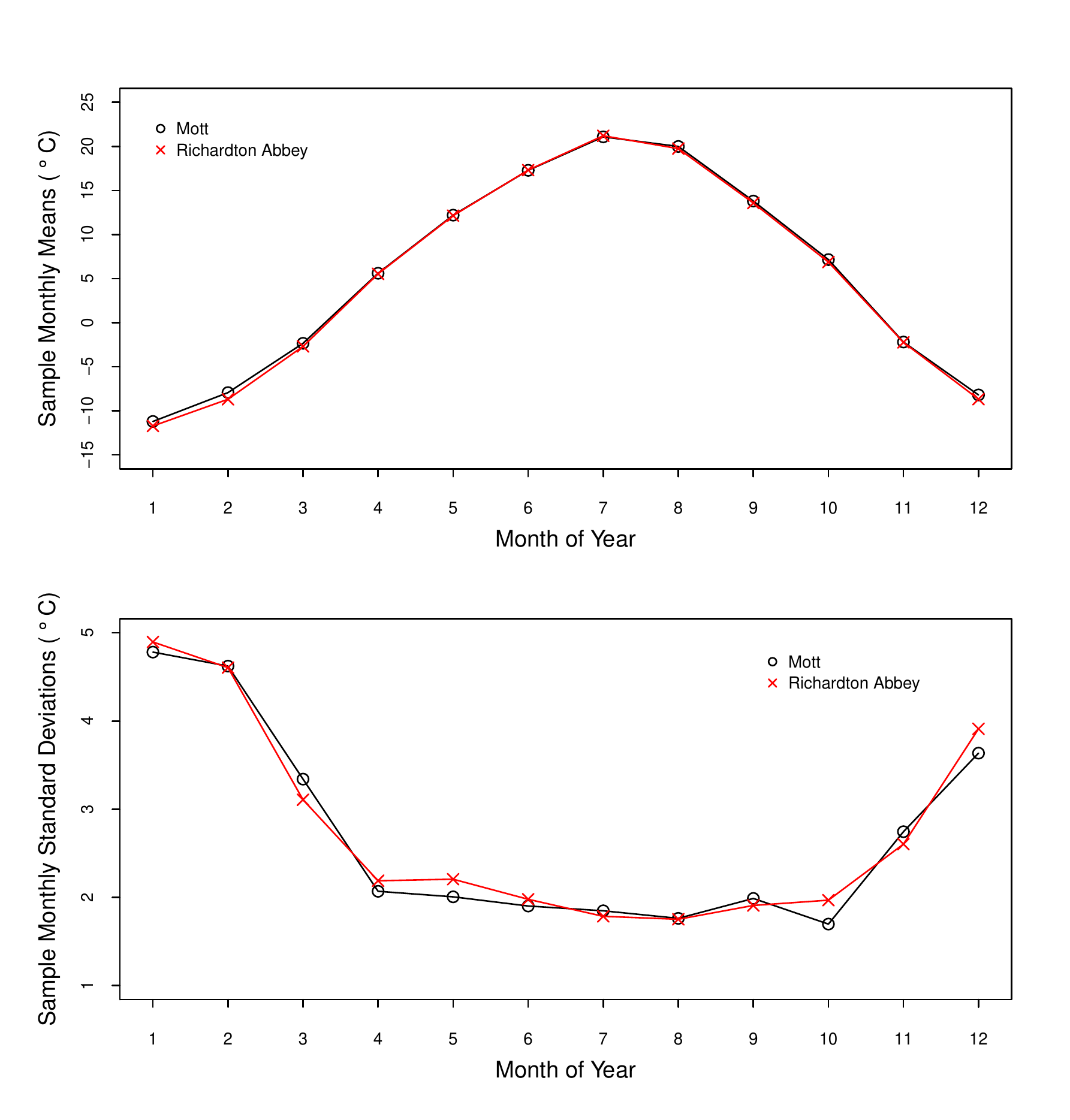}\\
	\caption{Monthly sample means (Top) and standard deviations (Bottom) for the Mott and Richardton-Abbey stations.}
	\label{NDspecs}
\end{figure}

A prominent seasonal mean cycle exists in the plotted data in Figure \ref{NDdata}.  In fact, the yearly range of the sample means exceeds $30^{\circ}$C: from a January minimum of less than -10$^\circ$C to a July maximum of more than $20.0^\circ$C.  This seasonal cycle makes it difficult to see small shifts, say on the order of a degree of two (the typical discontinuity magnitude induced by a changepoint), in the record.  These small changepoint shifts become critical in assessing long term changes in temperatures. Figure \ref{NDspecs} shows the monthly sample means and standard deviations of these two series.

Seasonality is also present in the variability of many climate series. The sample standard deviations in Figure \ref{NDspecs} show that winter temperatures are much more variable than summer temperatures; temperate zone examples exist where the January standard deviations of winter temperatures can be some five times larger than summer standard deviations \citep{Lund_etal_1995}.  This same reference also shows that a stationary model modulated for periodicities in mean and variance (or equivalently, the standard deviation - the square root of the variance) adequately describes many periodic climate series $\{ X_t \}$:
\begin{equation}
	X_{nT +\nu} = \mu_\nu + \sigma_\nu S_{nT+\nu}.
	\label{tsmodel}
\end{equation}
\noindent Here, $X_{nT+\nu}$ is the series observation during the $\nu$th season of the $n$th cycle, $T$ is a known period ($T=12$ for monthly data), $\{ S_t \}$ is a zero mean unit variance stationary time series in time $t$ ($t=nT+\nu$), and $\sigma_{\nu}$ is the process standard deviation during season $\nu$. Trends and changepoint features are neglected (for the moment) in the above model.

Seasonal features make changepoint analyses difficult if not taken into account.  Elaborating, in time series plots, it can be difficult to visually discern the impact of a changepoint in a temperature series, which typically shifts a series only by a degree or two, when the series has a seasonal cycle of 40 degrees. In a multiple changepoint analysis of a daily series, the methods may flag many spurious changepoints within a year in an attempt to follow the seasonal mean cycle should the seasonal cycle be ignored in the modeling procedure. 

\subsection{Autocorrelation}

Temporal autocorrelation, which measures the tendency for adjacent observations in time to be similar/dissimilar, is often present in climate data. Autocorrelation is typically positive in temperature and other climate series; for example, hot and cold periods often cluster in runs of days or months.  Like seasonality, autocorrelation hinders detection of mean shifts.  This is because long runs of above/below normal temperatures, attributable to correlation, can be mistaken as a shift.

The correlation between $X_t$ and $X_{t+h}$ is defined as 
\[
\mbox{Corr}(X_t, X_{t+h})= \frac
{\mbox{Cov}(X_t, X_{t+h})}
{\mbox{Var}(X_t)^{1/2} \mbox{Var}(X_{t+h})^{1/2}},
\]
\noindent where $\mbox{Cov}(X_t, X_{t+h})= E[ X_t X_{t+h} ] - E[ X_t ]E[ X_{t+h} ]$.  In terms of (\ref{tsmodel}), $\mbox{Corr}(X_t, X_{t+h}) = \mbox{Corr}(S_t, S_{t+h})$. A clarification here: data should be deseasonalized (i.e., subtracting the seasonal mean cycle) before correlation calculations, a practice followed here.  This is because seasonal mean cycles are deemed fixed and not a contributor to variability; however, some authors view the seasonal cycle as a part of annual variability. For more concreteness, our estimates of the seasonal mean and variance during season $\nu$ are, respectively, 
\[
\hat{\mu}_\nu = 
d^{-1}\sum_{n=0}^{d-1}X_{nT+\nu} = \hat{E}[X_{nT+\nu}], 
\quad 
\hat{\sigma}_\nu^2=
\frac{\sum_{n=0}^{d-1} (X_{nT+\nu}-\hat{\mu}_\nu)^2}{d-1},
\]
and our estimate of the lag $h$ correlation in $\{ S_t \}$ is
\[
\widehat{ \mbox{Corr} }
(S_t, S_{t+h})=
\frac{1}{dT}\sum_{\ell=1}^{dT-h} \hat{S}_t \hat{S}_{t+h}. 
\]
Here, $d$ is the number of complete cycles of data (we assume that no partial years of data are observed to avoid trite work) and hats indicate estimates of quantities.  Note that the first cycle of data is indexed by $n=0$ and the last by $n=d-1$.  Some authors use $d$ in place of $d-1$ in the definition of $\hat{\sigma}_\nu$; others use $dT-h$ in place of $dT$ in the definition of $\widehat{\mbox{Corr}}(S_t, S_{t+h})$.

Figure \ref{NDspecs} shows sample correlations from the monthly Mott and Richardton-Abby stations along with $95\%$ pointwise confidence bounds for zero correlation (white noise).  Notice that significant non-zero correlation exists at both stations.  

\begin{figure}[t] \centering 
	\noindent\includegraphics[width=\textwidth,angle=0]{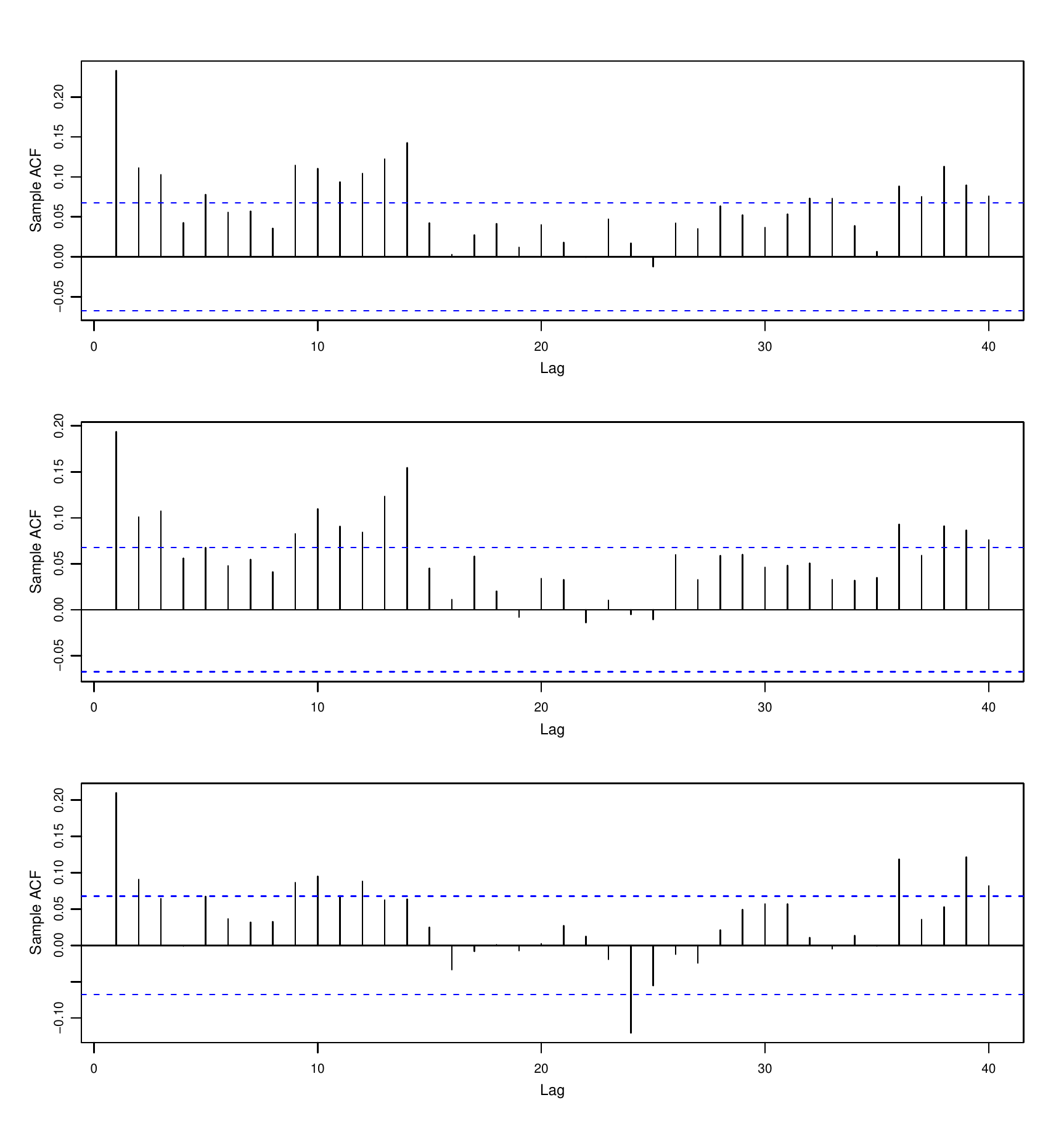} \\ 
	\caption{Sample correlations at the Mott (Top), Richardton-Abbey (Middle), and the Mott minus Richardton-Abby (Bottom) series after a seasonal standardization for each series over the first 40 months.   While the auto correlations in the two individual series are similar, correlation does not vanish in the target minus reference series in the bottom plot.} 
	\label{NDCorr} 
\end{figure}

Statistical methods for changepoint detection often degrade when correlation is present.  Indeed, an example is shown below where a changepoint declaration is repealed once correlation is taken into account. A key aspect of this paper is how to deal with cases where both correlation and mean shift changepoints are plausible.

\subsection{Target-Reference Comparisons}

Climate homogenization seeks to adjust time series for artificial features only, such as station relocations and instrumentation changes.  Natural/anthropogenic changepoints occasionally exist in series and are generally thought to be part of the record that should be retained.  To facilitate this, climatologists often make target-reference comparisons.  A reference series is a record of like data collected geographically near the target series (that hopefully experiences similar weather).   When one series is subtracted from the other, the target minus reference subtraction serves to remove natural fluctuations, especially if the target and reference series experience similar weather.  This subtraction reduces or altogether eliminates seasonal cycles and trends (more on trends below),  helping to highlight changepoints in the record.  Of course, any changepoint in either the target or reference series becomes a changepoint in the differenced series, so some negatives are incurred in the comparison. 

The bottom plot in Figure \ref{NDdata} shows the Mott series subtracted from the Richardton-Abby series.  Observe that the seasonal cycle has lessened, if not altogether disappeared.  If the target-reference comparison is good, any trend experienced by the target series should also be experienced by the reference series and removed (or greatly reduced) in the subtraction.  The lower plot in Figure \ref{NDCorr} shows sample correlations from the Mott minus Richardton-Abby stations along with $95\%$ pointwise confidence bounds for zero correlation (white noise).  These are correlations computed after a sample monthly mean has been subtracted from the series and the series has been further divided by a monthly sample standard deviation to put all temperatures on the same mean zero unit variance scale.  Notice that significant non-zero correlation exists at both stations.   Unfortunately, a target minus reference subtraction will not generally eliminate correlations; in fact, they often serve to increase them.

Since the statistical methods to conduct a changepoint analysis on the target series alone or the target minus reference series are the same, this point is essentially moot in the rest of this paper; nonetheless, it's practical implications are profound.  The authors are strong proponents of using target minus reference comparisons.  We refer the reader to \cite{Menne_Williams_2005, Menne_Williams_2009} for more on target-reference comparisons.  Some modern methods use multiple references series, sometime as many as forty \cite{Menne_Williams_2005, Menne_Williams_2009}.

\subsection{Trends} 

Many climate series have long term trends.  For example, in the Mott series in Figure \ref{NDdata}, a long term linear trend of $0.858 ^\circ C$/Century is estimated (computed neglecting any changepoints).  Of course, many temperature series exhibit recent warming, and many other climatic series have substantial trends as well. Trend features will be important to account for in changepoint analyses: a multiple changepoint procedure applied to a series with a trend that is ignored in the modeling procedure will typically flag multiple mean shifts in an attempt to follow the trend. 

While here, trends and seasonal means are features in the series mean $E[ X_t ]$.  Since a changepoint analysis typically tries to find abrupt changes in $E[ X_t ]$, it is imperative that these features should be competently described in the modeling scheme.  Unfortunately, even for single changepoint tests, the best test statistic to use will change depending on the true form of $E[X_t]$.  While one goal of this article is to deconvolve this issue for the reader, the limit distribution (and test percentiles) for a test with a long term trend are substantially different from those without a trend.

\subsection{Normality}

Climate time series may or may not be Gaussian (normal).  A series is Gaussian if its marginal distributions come from the normal distributional family.  Series that are averaged --- like monthly or annual series that are obtained by averaging daily data --- are often very close to normal by the central limit effect \citep{kwak2017central}. This can be visually checked by plotting a histogram of the series; normal data should have a unimodal symmetric histogram.  A Q-Q (quantile-quantile) plot gives a graphical check for normality where points lying on the diagonal indicates the data is well described by a Gaussian model.  A commonly used and powerful non-parametric statistical test for normality is the Shapiro-Wilks test.  The $p$-value for the Shapiro-Wilks test is $0.734$, reinforcing that normality is quite plausible for the Mott series (see \cite{yazici2007comparison} for more on normality tests). We do however need to stress that the Gaussian assumption is placed on the residuals (from $E[ X_t ]$) thus if there is trend, seasonality or changepoints within the data these must before removed before testing.

Some climate series are decisively non-Gaussian.   Examples include discrete categorical series of cloud cover, ordered from zero (clear sky) to ten (complete overcast), zero-one series describing an on/off phenomenon, and series whose marginal distributions are skewed (such as annual precipitation).  Averaging tends to induce normality.  For example, while the monthly averaging of daily data above rendered the Mott series essentially Gaussian, daily data is often skewed and non-normal.  In fact, daily temperatures at temperate zone stations often have a heavy left tail (are skewed), especially in winter; see \cite{lund2006parsimonious} for an example. 

\section{Time Series Models}

The classical decomposition of a time series $\{ X_t \}$ has the form
\begin{equation}
	\label{eq:MODEL}
	X_t = \mu_t + s_t + \epsilon_t,
\end{equation}
\noindent where $\{ X_t \}$ is the observed series, $\{ \mu_t \}$ is a long-term trend, $\{ s_t \}$ is a deterministic seasonal cycle having known period $T$, and $\{ \epsilon_t \}$ is zero mean random error that is possibly correlated in time.  Most changepoint scenarios for univariate series can be worked into the form in (\ref{eq:MODEL}).  The seasonal cycle $\{ s_t \}$ is periodic in that $s_{t+T}= s_t$ for all times $t$.  When the parametrization for $\{ \mu_t \}$ contains a location parameter, one typically assumes that $\sum_{t=1}^T s_t = 0$ so that regression parameters are statistically identifiable.  This is the so-called classical decomposition model in \cite{Brockwell_Davis_1991}.

For a simple example, suppose that one is examining an annual series for possible multiple mean shifts, permitting a possible background linear time trend.  Then $T=1$, $s_t \equiv 0$, and $\mu_t = \beta_0 + \beta_1 t$ for a location parameter $\beta_0$ and trend parameter $\beta_1$.  The regression model can be written as
\begin{equation}
	\label{eq:slr}
	X_t = \beta_0 + \beta_1 t + \delta_t + \epsilon_t,
\end{equation}
\noindent where the mean shift changepoint component $\{ \delta_t \}$ has the form
\[ 
\delta_t = 
\left\{ 
\begin{array}{l l}
	\Delta_1=0,   & \quad \tau_0 \leq  t < \tau_1,     \\
	\Delta_2,     & \quad \tau_1 \leq t  < \tau_2,     \\
	\vdots                                             \\
	\Delta_{m+1}, & \quad \tau_m \leq t  <  N.         \\ 
\end{array} \right. 
\]
\noindent The above setup takes data observed at the times $1, 2, \ldots N$ and allows for $m$ mean shift changepoints occurring at the times $\tau_1 < \tau_2 < \cdots < \tau_m$; the changepoint count $m$ and their occurrence times $\tau_1, \ldots, \tau_m$ are all unknown.  If the location parameter $\beta_0$ is omitted from the long-term trend expression, one need not require that $\Delta_1=0$.

A prominent seasonal cycle $\{ s_t \}$ exists in most temperate zone series; in general, the ``extra variation'' induced by the seasonal cycle makes changepoints harder to see and detect. The random errors $\{\epsilon_t \}$ are generally correlated with climate data.  Positive autocorrelation reduces the effective number of independent observations, making it harder to detect changepoints.

In what follows, our primary focus lies with the detection of mean changes in a series --- the so called mean shift problem.  This problem keeps the autocovariance structure of $\{ \epsilon_t \}$ constant across the entire series.  Changepoint methods exist for autocovariance changes \citep{Davis_etal_2006}, or even changes in the marginal distributions of the series \citep{Gallagher_etal_2012}, but the major focus on the climate literature to date has been on mean shifts.  Toward this, the mean shift changepoints here shift all subsequent series values by the same amount --- shifts are not seasonal. While the methods here could be modified to induce a seasonal change, this extension is not considered here.

When $T>1$, such as for a monthly series, it is convenient to express the regression model in a periodic form:
\begin{equation}
	\label{eq:sMODEL}
	X_{nT+\nu} = \mu_{nT+\nu} + s_\nu + \delta_{nT+\nu} + \epsilon_{nT+\nu},
\end{equation}
\noindent where the season $\nu \in \{ 1, 2, \ldots , T \}$ and $n$ indicates the cycle number corresponding to time $nT+\nu$.  For example, a regression model allowing for a different linear trend between all consecutive changepoint times has form
\[ 
\mu_t = 
\left\{ 
\begin{array}{l l}
	\beta_1 + \alpha_1 t,       & \quad \tau_0 \leq  t  < \tau_1,     \\
	\beta_2 + \alpha_2 t ,      & \quad \tau_1 \leq  t  < \tau_2,     \\
	\vdots                                             \\
	\beta_{m+1}+ \alpha_{m+1}t, & \quad \tau_m \leq t   < N.           \\ 
\end{array} \right. 
\]

The time series component $\{ \epsilon_t \}$ is typically assumed to be stationary when $T=1$, or periodically stationary when $T > 1$.  A good and flexible model class for stationary series are the autoregressive (AR) series \citep{Brockwell_Davis_1991}.  A $p$th order zero mean autoregression is uniquely characterized by the $p$-th order linear difference equation
\[ 
\epsilon_t = \phi_1 \epsilon_{t-1} + \ldots + \phi_p \epsilon_{t-p} + Z_t, 
\]
\noindent where $\phi_1, \ldots, \phi_p$ are the autoregressive coefficients and $\{ Z_t \}$ is a zero mean white noise sequence with variance $\sigma^2$.  When $T>1$, AR models are replaced with periodic AR models (PAR):
\[
\epsilon_{nT+\nu} = \phi_1(\nu) \epsilon_{nT+\nu-1} + \ldots + 
\phi_p(\nu) \epsilon_{nT+\nu-p} + Z_{nT+\nu}, 
\]
\noindent where $\phi_1(\nu), \ldots, \phi_p(\nu)$ are the autoregressive parameters during season $\nu$ and $\{ Z_{nT+\nu} \}$ is periodic white noise having a periodic variance $\mbox{Var}(Z_{nT+\nu})=\sigma^2_{\nu}$. PAR models can have a large number of parameters and are generally non-parsimonious.  For example, a PAR(3) model for a monthly series has 36 AR parameters and 12 more white noise parameters.  Our later examples will exhibit some techniques to reduce these parameter counts.

\section{Single Changepoint Detection}

\subsection{A single mean shift}

The simplest changepoint test discerns whether a series has no mean shifts (the null hypothesis) against the alternative hypothesis that there exists precisely one mean shift occurring at some unknown time.  These are the so-called at most one changepoint (AMOC) methods.  For the moment, assume that no long term trends exist in the series.  Almost all AMOC mean shift changepoint methods compare sample means of the series before and after all possible candidate changepoint times.  That is, they compare differences between $k^{-1}\sum_{t=1}^k X_t$ and $(N-k)^{-1}\sum_{t=k+1}^N X_t$ for each admissible changepoint time $k$, selecting the $k$ where this difference is statistically maximal as the changepoint time estimate.

Formalizing this, suppose first that $\{ \epsilon_t \}$ is independent and identically distributed (IID) with zero mean and variance $\sigma^2$.  A scaled version of these sample mean differences that takes into account the differing number of observations in the two segments is the cumulative sum statistic having a changepoint at time $k$:
\[
\mbox{CUSUM}_X(k) = \frac{1}{\hat{\sigma} \sqrt{N}} 
\left[ \sum_{t=1}^k X_t - \frac{k}{N}\sum_{t=1}^N X_t \right],
\]
where
\[
\hat{\sigma}^2= 
\frac{\sum_{t=1}^N (X_t -\bar{X})^2 }{N-1} 
\]
is the no changepoint null hypothesis estimate of the series' variance and $\bar{X}=N^{-1}\sum_{t=1}^N X_t$ is the overall sample mean.  One takes the argument $k$ that maximizes $|\mbox{CUSUM}_X(k)|$ as the estimated changepoint time.  

\vspace{.12in} \noindent {\bf Pitfall 1:} Many past climate changepoint authors examine a ``maximum statistic'' akin to $D^*=\max_{2 \leq k \leq N} |\mbox{CUSUM}_X(k)|$ to check for a single changepoint. Where the maximum occurs is estimated as the time of the changepoint.  While this is fine, incorrect null hypothesis distribution percentiles abound in the climate literature, often producing unjustifiable conclusions \citep{Lund_Reeves_2002, Robbins_etal_JTSA}.  When a changepoint is known to occur at time $k$, $\mbox{CUSUM}_X(k)$ should be used as the test statistic.   Scaling $\mbox{CUSUM}_X(k)$ to a $z$, $t$, or even $F$ distribution is used to make conclusions.  When the time of the changepoint is unknown, the maximum statistic $D^*$ is used. The correct null hypothesis percentiles for $D^*$ must account for the many times $k$ where the maximum could arise --- these percentiles are much larger than those for a fixed $k$.  The correct asymptotic quantification of AMOC changepoint statistics is often unwieldy as the scenario is not readily scaleable to an extreme value distribution.  Indeed, $\mbox{CUSUM}_X(k)$ are highly correlated in $k$ (they are not IID).  The distribution of AMOC tests often converges to the supremum of some Gaussian process.  The reader is referred to \cite{Csorgo_1997_LimitThm, MacNeil_l1974_BM} for historical technical development.

\vspace{.12in} \noindent {\bf Best Practice 1:} Several legitimate statistics can be used to detect a single mean shift changepoint.  One test with great detection power uses a sum of squared CUSUM statistics to assess whether a changepoint is present:
\[
\mbox{SCUSUM}=\sum_{k=1}^N \mbox{CUSUM}_X^2(k).
\]
The time of the changepoint is still estimated as the location $k \geq 2$ that maximizes $|\mbox{CUSUM}_X(k)|$. This test won the single changepoint comparison competition in \cite{Shi_etal_2021}, originates from \cite{Kirch_2006_CUSUM}, has good false detection properties and superior detection power.

Under the null hypothesis of no changepoints, the asymptotic distribution of the $\mbox{SCUSUM}$ test converges to that of $\int_0^1 B^2(t)dt$, the integrated square of a standard Brownian Bridge stochastic process \citep{Shi_etal_2021}.  The null hypothesis percentiles of this distribution are presented in Table \ref{tab:scusum_table} for convenience and are extracted from \cite{Shi_etal_2021}.  While the SCUSUM test does not appear to be frequently used in today's climate literature, summing CUSUM statistics over all times increases detection power.  As such, we recommend this test in single changepoint analyses.  Additional discussion is contained in \cite{Shi_etal_2021}.

\begin{table}[H]
	\caption{Critical Values for SCUSUM Statistics}
	\begin{center}
		\begin{tabular}{ c  c }
			\hline
			Percentile    &    Critical Value  \\
			\hline
			90.0\%      & 0.3473046       \\
			95.0\%      & 0.4613744       \\
			97.5\%      & 0.5806168       \\
			99.0\%      & 0.7434348       \\
			\hline
		\end{tabular}
		\label{tab:scusum_table}
	\end{center}
\end{table}

\subsection{Autocorrelation}
We now move to AMOC tests in correlated data.  A significant body of statistical research modifies the limit theory for IID data to account for autocorrelation \citep{Robbins_etal_JTSA, Shi_etal_2021}. Much of this literature takes the following flavor.  With the SCUSUM test above (and other AMOC tests), simply replace $\hat{\sigma}$ with an estimate of the long-run variance parameter $\tau$ defined by
\[
\tau^2 = \lim_{N \rightarrow \infty} N \mbox{Var}
\left( N^{-1} \sum_{t=1}^N X_t \right).
\]
Then most asymptotic limit laws apply with this simple modification.  For example, should $\{ X_t \}$ be a short memory covariance stationary series with lag-$h$ covariance $\gamma(h)=\mbox{Cov}(X_t, X_{t+h})$ (such as an ARMA model), then 
\[
\tau^2 = \gamma(0) + 2 \sum_{h=1}^\infty \gamma(h).
\]
These tests should not be applied to long-memory series where $\tau^2$ can be infinite. In practice, it is not clear how to best estimate $\tau^2$, which is the notorious spectral density at frequency zero.  A recent statistical reference on this topic is \cite{Spectral-1984-Zeros}.

In some asymptotic tests, convergence to the limit law can be slow, making application to even a century of annual data questionable. A preferable way to handle correlation involves pre-whitening techniques.  Statistical reference for pre-whitening are \cite{Robbins_etal_JTSA, Gallagher-2022-Autocovariance}. To account for correlation in an AMOC changepoint analysis, pre-whitening first fits a $p$-th order autoregressive (AR($p$)) model to the series (this assumes annual or non-periodic data). This fit is conducted under the null hypothesis of no changepoints and is easily accomplished with many standard time series analysis packages. This procedure yields estimates of the autoregressive parameters $\phi_1, \ldots, \phi_p$ and the white noise variance $\sigma^2$. Next, the one-step-ahead predictions
\[ 
\hat{X}_{t+1}= \hat{\phi}_1 X_t + \ldots + \hat{\phi}_p X_{t-p+1}, 
\quad t \geq p,
\]
\noindent are calculated with $\hat{\phi}_j$ replacing $\phi_j$ and the one-step-ahead prediction errors $Y_t = X_t -\hat{X}_t$ are formed.  When the AR($p$) parameters are known, the one-step-ahead prediction errors $\{ Y_t \}$ are independent.  Using estimated AR parameters leaves the $\{ Y_t \}$ slightly dependent, but this dependence is usually negligible. The series $\{ Y_t \}$ is called the pre-whitened series.
To compute the startup values $\hat{X}_1, \ldots, \hat{X}_p$, one uses the time series prediction equations; see Chapter 3 of \cite{Brockwell_Davis_1991} for details.

Next, one simply applies the SCUSUM (or some other AMOC) test to the pre-whitened $\{ Y_t \}$ using the percentiles for IID errors to make conclusions.  \cite{Robbins_etal_JTSA} proves that this procedure is statistically valid asymptotically.  More importantly, this paper shows that the limit laws typically ``kick in more quickly'' than asymptotic laws that replace $\hat{\sigma}$ with $\hat{\tau}$.

While pre-whitening adds another layer to the analysis, our next pitfall notes the importance of taking correlation into account.

\vspace{.12in} \noindent {\bf Pitfall 2:} Ignoring positive correlation in a series will often produce spurious changepoint conclusions.  In fact, series that are heavily positively correlated tend to have long sojourns above and below the long-term mean of the series, inducing the appearance of a changepoint.  Ignoring correlation may induce the spurious conclusion that a changepoint exists when in truth it does not.

\vspace{.12in} \noindent {\bf Best Practice 2:} Pre-whiten any autocorrelated series before applying AMOC IID changepoint tests.  As shown below, dubious conclusions can arise when autocorrelation is ignored. A general theme for AMOC tests with positively correlated data, which entail the majority of climate cases, is clear:  one risks concluding that a changepoint exits when in truth it does not when positive correlation is ignored. The situation reverses itself should negatively correlated data be encountered.

\subsection{An Example}

\begin{figure}
	\centering
	\includegraphics[scale=1]{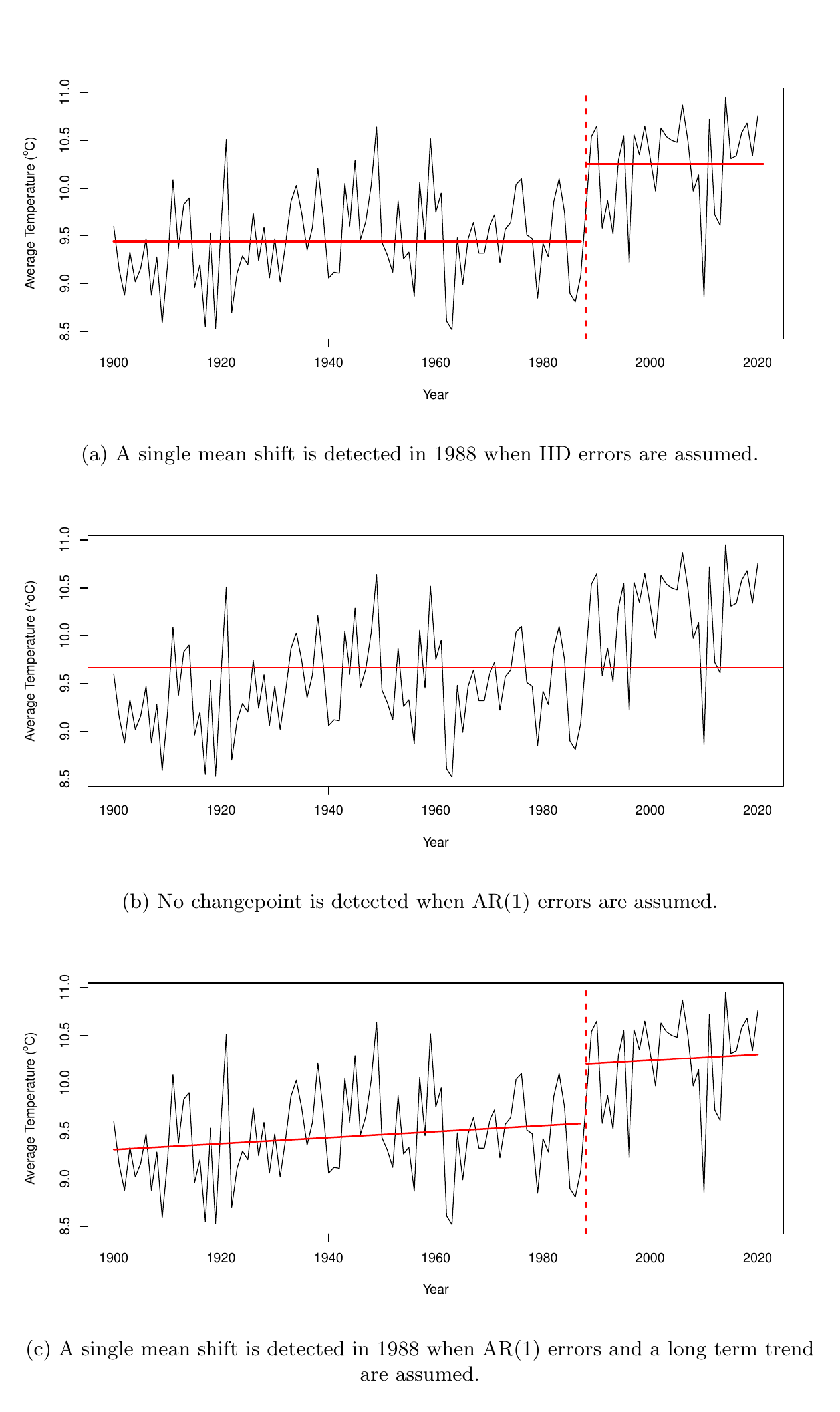}
	\caption{Single changepoint tests flag a mean shift in 1988.}
	\label{fig:SCPT-plot}
\end{figure}

We now examine the annual Central England temperature (CET) series from 1900-2020 with a single changepoint test. The CET record was provided by the UK Met Office at \url{https://www.metoffice.gov.uk/hadobs/hadcet/}. For a multiple changepoint analysis of the entire series dating back to the 1600s, see \cite{Shi-2022-CET}.  Figure \ref{fig:SCPT-plot} plots this series against several single changepoint configurations explored below.  Conclusions are shown below to be heavily dependent on the assumptions made.

As a first step, we examine the series for a single mean shift assuming IID errors.  The $\text{CUSUM}(k)$ statistic is maximized at $k=1988$ and the $\text{SCUSUM}$ statistic is $3.577$.  Comparing to the $95^{\text{th}}$ percentile of $\text{SCUSUM}$ statistic, which is is $0.4614$, one concludes that a mean shift exists with confidence at least $95\%$ (in fact, the $p$-value of erroneously rejecting a no changepoint null hypothesis is zero to about six decimal places).  The estimated mean configuration is plotted against the series in the top panel of Figure \ref{fig:SCPT-plot}.

We now rerun the single mean shift test allowing for autocorrelated errors, this time using a simple AR(1) structure for the model errors. An SCUSUM test was applied to the pre-whitened AR(1) one-step-ahead prediction errors and gives $\text{SCUSUM}_Z=0.1799$, which is well below the $0.4614$ threshold needed to declare statistical significance with $95\%$ confidence (the $p$-value for this test is $0.31$). This essentially repeals the 1988 mean shift changepoint declaration made in the above paragraph. Here, the estimated AR(1) correlation between consecutive series observations is $\hat{\phi}=0.425$, which is not extreme autocorrelation.  The conflicting conclusions illustrate why one needs to be careful to allow for correlation in changepoint tests when correlation is present; neglecting to account for positive correlation often leads to an overestimation of the number of changepoints.

\begin{table}[H]
	\caption{Single Changepoint Tests for the Central England Series}
	\begin{center}
		\begin{tabular}{ c c c }
			\hline
			Model Assumptions         & Test   & $p$-value  \\
			\hline
			Mean shift  + IID errors   & $\text{SCUSUM}$ & $\leq 10^-6$ \\
			Mean shift  + AR(1) errors & $\text{SCUSUM}_Z$ & $0.31$ \\
			Fixed trend + AR(1) errors & $\text{CUSUM}_D$ & $0.039$ \\
			\hline
		\end{tabular}
		\label{tab:scpt_test}
	\end{center}
\end{table}

\subsection{Trends}
As previously mentioned, trends can also influence changepoint conclusions.  In particular, one should not apply a changepoint test to data with a trend without accounting for the trend.  For example, should the linear trend $\mu_t = \beta_1 t$ exist in the CET series but not modeled, then an AMOC test tends to signal a single changepoint in the center of the record with a positive mean shift when $\beta_1 > 0$, and flag a negative mean shift in the center of the record when $\beta_1 < 0$. The methods are simply rejecting that the mean is constant (which is why some authors use changepoint tests as a check for a constant mean).  When a seasonal cycle exists in the data, the situation becomes even more nebulous, with multiple changepoint techniques flagging multiple changes in an attempt to ``follow'' the seasonal mean and long term trend.  In short, changepoint techniques are not robust to changes in $\mu_t=E[X_t]$.  Unfortunately, in changepoint analyses, each different form of $m_t$ requires a different set of null hypothesis percentiles. For example, in the case of a simple mean shift where $\mu_t=\beta_0$, the $95^{\text{th}}$ percentile of the CUSUM test is $1.358$; when there is a linear trend $\mu_t=\beta_0 + \beta_1 t$, the $95^{\text{th}}$ CUSUM percentile becomes $0.902$.

\vspace{.12in} \noindent {\bf Pitfall 3:}  Applying an AMOC changepoint test to series with trends or seasonality that does not account for the trend or seasonality can result in spurious changepoint declarations.  Here, the methods are simply declaring  that the series does not have a constant mean.   

\vspace{.12in} \noindent {\bf Best Practice 3:}   Account for all features in the mean of a series. If in doubt, allow for a trend and/or seasonality and use the statistical methods to distinguish which features are present in the series.

\subsection{CET Example Rejoinder}

As an example, we return to the 1900-2020 CET series.  Global warming posits a slow temperature increase; as such, an AMOC analyses with the linear trend $\mu_t = \beta_0 + \beta_1 t$ is explored.  AR(1) errors are also allowed in the procedure.  An AMOC CUSUM-type mean shift test for IID errors is developed in \citet{Gallagher_etal_2012} (we are unaware of anyone studying SCUSUM tests in the linear trend setting). This test is denoted by $\mbox{CUSUM}_D$.  Estimating the linear trend and AR(1) specifications under the null hypothesis of no changepoints provides $\hat{\beta}_0=9.1^\circ$ C, $\hat{\beta}_1=-0.0092^\circ$ C/year, and $\hat{\phi}=0.194$.


One needs to be careful to account for the trend when pre-whitening this series.  Specifically, our estimated one-step-ahead prediction with AR(1) errors is 
\begin{equation}
	\label{blp}
	\hat{X}_t = 
	\widehat{\mu_t} + \hat{\phi} (X_{t-1} - \widehat{\mu_{t-1}})=
	\hat{\beta}_0 + \hat{\beta}_1 t + \hat{\phi} 
	\left[ X_{t-1} - \hat{\beta}_0 - \hat{\beta}_1 (t-1) \right], 
\end{equation}
for $t \geq 2$, with the start-up condition $\hat{X}_1= \hat{\beta}_0 + \hat{\beta}_1$.  The pre-whitened series is always $Y_t= X_t -\hat{X}_t$.

The $\mbox{CUSUM}_D$ test applied to $\{ Y_t \}$ gives a statistic of $0.929$, which is slightly above the $95^{\text{th}}$ percentile null hypothesis threshold of $0.9028$. The $p$-value for this test is $0.038$.  With $95\%$ confidence, the 1988 changepoint has come back.  The bottom panel in Figure \ref{fig:SCPT-plot} displays the fit to this data.  This configuration is the best fitting of our three models.  Overall, it is the most reliable fit as it takes into account both trends and autocorrelation.   See \cite{Shi-2022-CET} for a detailed analysis of the CET series.

Obviously, the assumptions made in changepoint analyses are extremely important and may influence conclusions.  While issues become even more complex in multiple changepoint settings, the topic of our next section, much of the AMOC intuition carries over to that setting.

\section{Multiple Changepoint Detection}

Many climate series have more than one changepoint.  United States climate series average a station move or gauge change once every 17 years, roughly \citep{Mitchell_1953}.  As in the AMOC case, multiple changepoint (MCPT) detection is also fraught with challenges and pitfalls, perhaps
more than the single changepoint case.  While MCPT analyses are less developed than AMOC tests, the problem is being actively researched in statistical settings.

Initially, AMOC techniques were extended to MCPT problems via binary segmentation methods \citep{Scott_1974_BS}. Binary segmentation examines the entire series first for a single changepoint with some AMOC test.  If a changepoint is found, the series is then split into two subsegments about the identified changepoint time and the two subsegments are further scrutinized with AMOC tests. The procedure continues iteratively until all subsegments are declared changepoint free.  We now know that binary segmentation is one of the poorer ways to handle multiple changepoint problems \citep{Shi_etal_2021}.  This point is further reinforced below.

Other approaches to the MCPT problem can be classified into distinct camps.  One camp examines recursive segmentation procedures that improve upon binary segmentation methods include wild binary segmentation \citep{Fryzlewicz_2014} and wild contrast maximization \citep{Cho-2020-WCM}.  These methods are computationally quick and often yield reasonable results.  Unfortunately, many of these techniques declare an excessive number of changepoints when the true number of changepoints is small \citep{Shi_etal_2021}, essentially rendering them unusable in climate cases where say two changepoints exist in a hundred year climate series.  

Another camp applies dynamic programming to MCPT problems.   Here, an objective function associated with the problem is optimized.  The segmented neighborhood algorithm of \cite{Auger_1989_SegNeighborhood} and the pruned exact linear time of \cite{Killick_2012_PELT} are two examples.  Dynamic programming techniques provide optimal (relative to the chosen objective function) changepoint configurations and can be computed very quickly.  Unfortunately, these techniques often make unrealistic assumptions (uncorrelated series or all model parameters must shift at every changepoint time) that make them unfeasible in some climate applications. Advances to these methods are currently being pursued in \cite{GD_PELT}. Model selection approaches such as \cite{Harchaoui_2010_TV} and \cite{Shen_2012_AdaLASSO} and scan statistics procedures based on moving sum statistics \citep{Kirch_2018_Mosum} also exist.  Other methods also exist --- changepoint research is a huge field and this list is not exhaustive.

Like the AMOC case, assumptions are crucial in MCPT analyses.  Many MCPT techniques assume IID $\{ \epsilon_t \}$, which is often unrealistic in climate applications.  MCPT techniques for independent $\{ \epsilon_t \}$ can give suboptimal answers for correlated series \citep{Davis_etal_2006, Lund_2012_GA, Chakar_2017_AR1}.  While one can still pre-whiten the series, estimation of the correlation structure and the multiple mean shift sizes and locations confound each other.  No null hypothesis suggests itself in the MCPT setting.   In the AMOC case, estimates of the series' correlation structure were computed under the null hypothesis of no changepoints and models containing no and one changepoints were statistically compared.  In the MCPT case, the number of changepoints in the null hypothesis is unclear.  Trends and seasonality further impede issues.

Penalized likelihood methods, another MCPT camp, attack the problem by minimizing a likelihood objective function that is penalized when the model contains too many changepoints.  Elaborating, statisticians often estimate model parameters via likelihood techniques.  Let $L(m; \tau_1, \ldots, \tau_m)$ denote the likelihood of the best time series model having $m$ changepoints at the times $1 < \tau_1 < \tau_2 < \cdots < \tau_m \leq N$.  Likelihoods for a Gaussian series $\{ X_t \}_{t=1}^N$ take the classical time series form
\[
L(m;\tau_1, \ldots, \tau_m)= (2 \pi )^{-N/2} 
\left( \prod_{t=1}^N V_t \right)^{-1/2} 
\exp{ \left[ - \sum_{t=1}^{N} \frac{(X_t -\hat{X}_t)^2}{V_t} \right]},
\]
where $\hat{X}_t$ is the best linear prediction of $X_t$ from past observations in (\ref{blp}) and $V_t=E[(X_t-\hat{X}_t)^2]$ is its unconditional mean squared error.

As the number of changepoints $m$ increases, the model fit improves: $L(m; \tau_1, \ldots, \tau_m)$ increases in $m$.  However, after a while, adding additional changepoints does not appreciably improve the likelihood.  This is where the penalty term comes in. The penalty for having $m$ changepoints at the times $\tau_1, \ldots, \tau_m$ is denoted by $P(m; \tau_1, \ldots , \tau_m)$ and grows as $m$ increases.   Penalized likelihood methods look to minimize the penalized objective function
\[
O(m; \tau_1, \ldots, \tau_m) = 
- 2 \ln (L(m; \tau_1, \ldots, \tau_m))+P(m; \tau_1, \ldots, \tau_m)
\]
over all feasible values of $m$ and $\tau_1, \ldots, \tau_m$.  When there are no changepoints ($m=0$), the penalty term is taken as zero. 

Development of appropriate penalty functions is a well-studied statistical problem.   Commonly used penalties for the mean shift problem with IID errors include the AIC, BIC, mBIC, and MDL.   Their formulas are
\begin{align}
	\mbox{AIC}  &:& P(m; \tau_1, \ldots, \tau_m)= 2(2 m + 2)    \nonumber \\
	\mbox{BIC}  &:& P(m; \tau_1, \ldots, \tau_m)= (2m+2)\ln(N) \nonumber \\
	\mbox{mBIC} &:& P(m; \tau_1, \ldots, \tau_m)= 3m\ln(N)+\sum_{i=1}^{m+1} \ln \left( \frac{\tau_i - \tau_{i-1}}{N}  \right )      \nonumber \\
	\mbox{MDL}  &:& P(m; \tau_1, \ldots, \tau_m)=\sum_{i=1}^{m+1} \ln (\tau_i - \tau_{i-1} )  +   2\ln(m) + 2\sum_{i=2}^{m} \ln ( \tau_i )        \label{eqn:pform}
\end{align}
The penalties above are for the changepoint configuration portion of the model only; should the time series structure change at each changepoint time, the above formulae require modifications. When the time series structure is constant across all series segments, the above formulae are appropriate.  While other penalties exist, these are the major penalties used in today's literature.   Note that the mBIC and MDL penalties depend on where the changepoints lie, but that the AIC and BIC penalties are simple multiples of the number of changepoints.  A detailed discussion of penalty performances is found in \cite{Shi_etal_2021}; however, AIC typically overestimates the true number of changepoints and is not used.  For a penalty that does not depend on the changepoint location times, BIC performs surprisingly well in a variety of settings \citep{Shi_etal_2021}.

In optimizing $O(m; \tau_1, \ldots , \tau_m)$, significant computational issues arise.   To compute $P(m; \tau_1, \ldots, \tau_m)$, an optimal time series model with $m$ changepoints at the times $\tau_1, \ldots, \tau_m$ needs to be fitted.   While this is a straightforward task for most time series packages, there are $2^{N-1}$ distinct changepoint configurations that need to be evaluated as candidates.  This total is immense for even $N$ as large as 100, making exhaustive identification of the best changepoint configuration a strenuous task. Authors have used genetic algorithms \citep{Davis_etal_2006, Li_Lund_2015} to overcome these difficulties.  Today, despite computational issues, penalized likelihood is considered the gold standard of MCPT problems.  Moreover, work is currently being pursued to bring rapid computation to the setting where the time series parameters are constant across all regimes \citep{GD_PELT}. 

\subsection{Binary Segmentation}

As the earliest invented and still widely used MCPT technique, binary segmentation's popularity rests on two ingredients: ease of interpretation and rapid computation.  Binary segmentation is a ``greedy algorithm'' that optimizes an objective function stagewise.  Such a procedure often does not find the globally optimal solution.   An attempted remedy to binary segmentation, wild binary segmentation \citep{Fryzlewicz_2014}, injects randomization into the changepoint search to avoid local optimums.  However, simulation studies in \citep{Lund_2020_WBS2} suggest that wild binary segmentation overestimates changepoint counts for IID model errors, and becomes dysfunctional in settings with correlated errors. Wild contrast maximization \citep{Cho-2020-WCM}, another improvement of wild binary segmentation designed for dependent processes, is capable of handling serial dependence.  While we will not discourage the user from using this technique, we also comment that it has not be fully vetted as of today. 

\vspace{.12in} \noindent {\bf Pitfall 4:  Using Ordinary Binary Segmentation in MCPT Problems}

\begin{figure}
	\centering
	\includegraphics[scale=0.8]{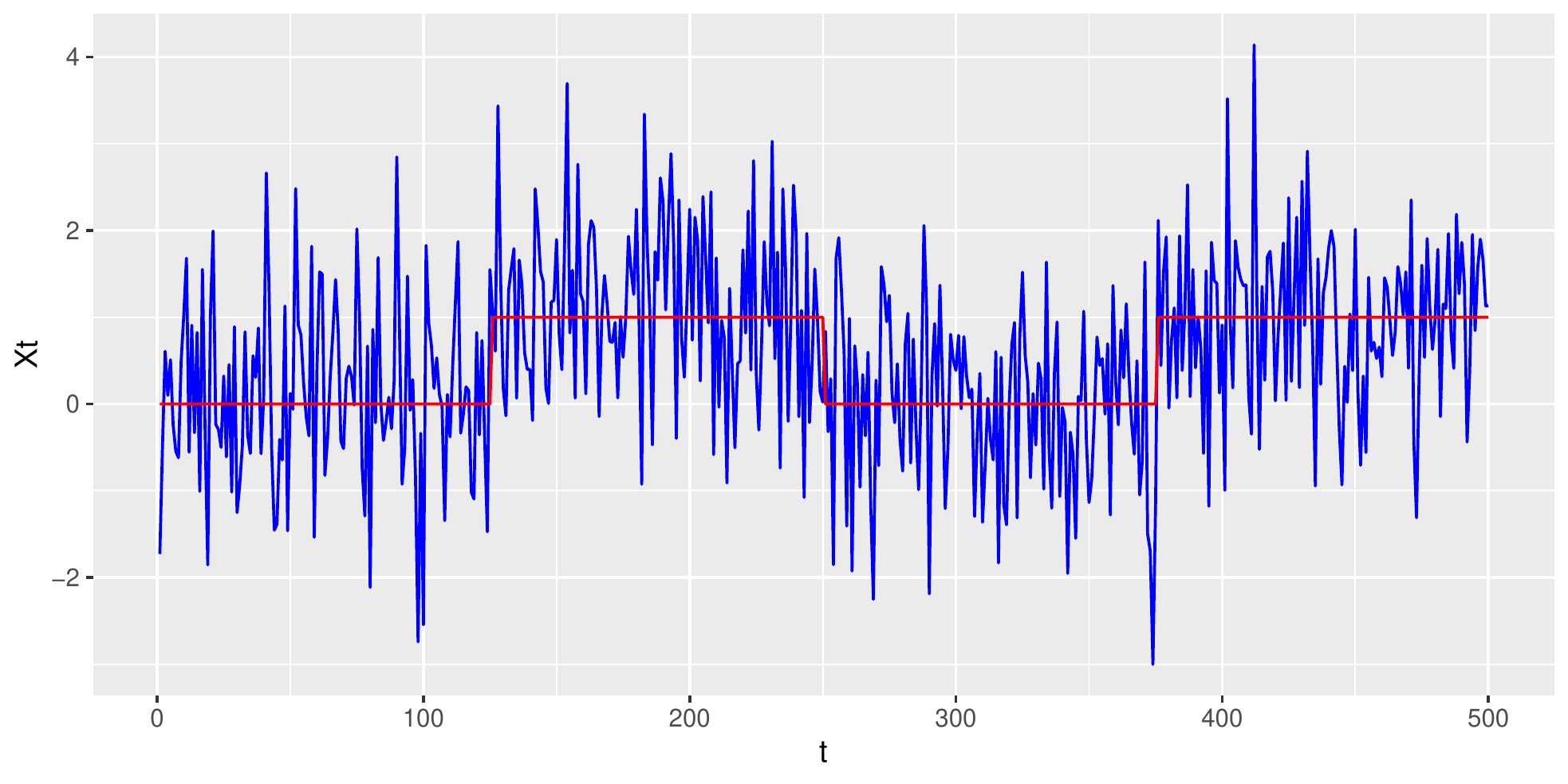}
	\caption{A series with three equally spaced mean shifts of unit size shifting the series in alternating directions. The regression errors are uncorrelated white noise with a unit variance.}
	\label{fig:Simulated_BS_CounterExample}
\end{figure}

Binary segmentation is generally an inferior MCPT problem approach, regardless of assumptions.  Unfortunately, binary segmentation is used in many engineering, computer science, and climate applications. To illustrate binary segmentation pitfalls, a simulation was constructed.  Here, Gaussian series of length $500$ were simulated with white noise errors. Three equally spaced changepoints were inserted shifting the series by a unit length in alternating directions. This partitions the series into four equal length segments of 125 points each; Figure \ref{fig:Simulated_BS_CounterExample} displays a sample generated series. 

\begin{figure}
	\centering
	\includegraphics[scale=0.8]{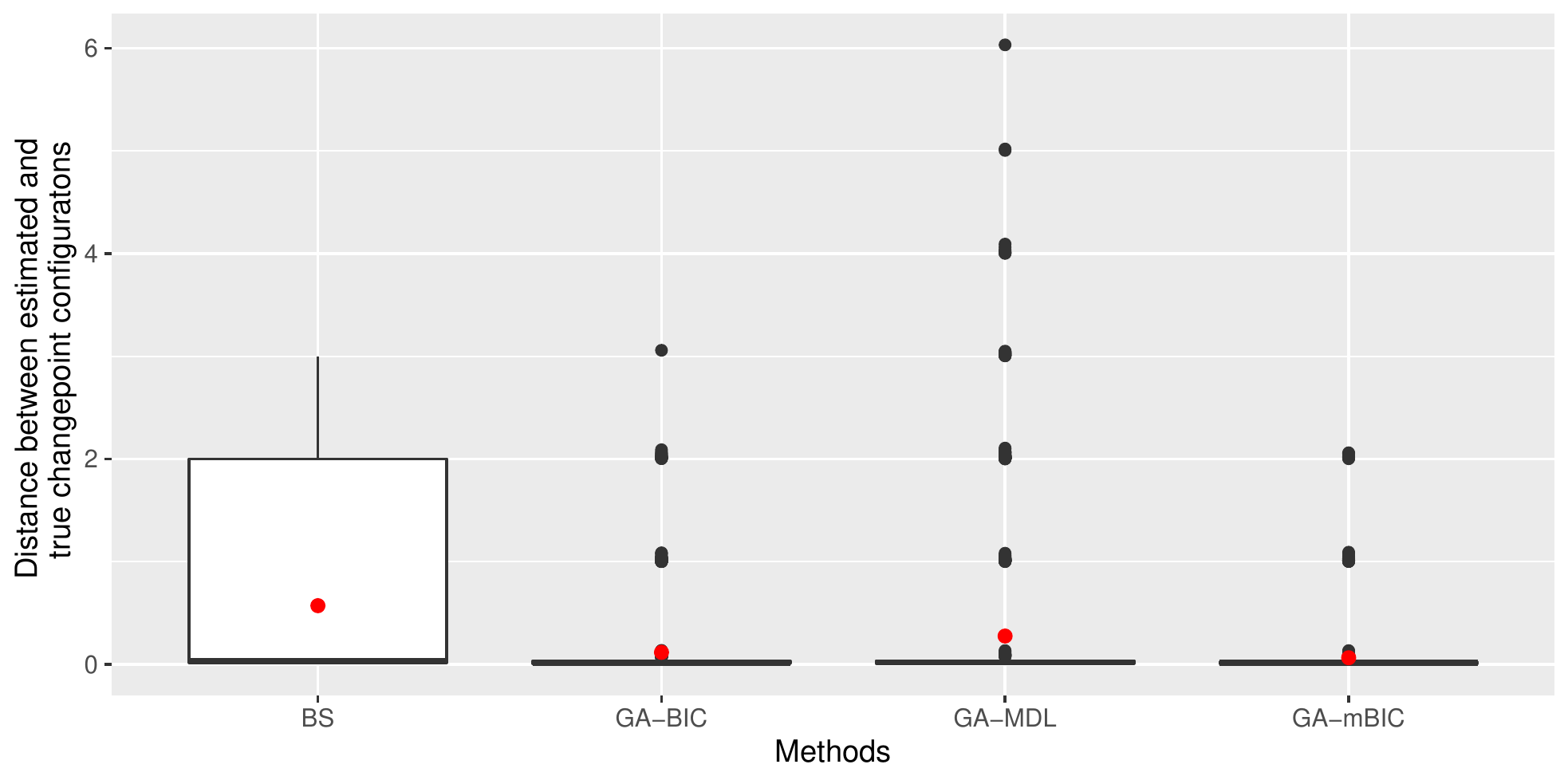}
	\caption{Binary segmentation and penalized likelihood methods compared.  The biggest errors occur with binary segmentation. Note $95\%$ threshold is used for binary segmentation and BIC, MDL and mBIC penalized likelihoods are optimized by the genetic algorithm (GA).}
	\label{fig:BS_CounterExample_Boxplots}
\end{figure}

We randomly generated 100 such series and applied several different changepoint methods. The estimated changepoint configurations were compared to the true changepoint configuration with the distance metric in  \cite{Shi_etal_2021}).  This distance incorporates both $m$ and the changepoint locations $\tau_1, \ldots, \tau_m$.  Smaller distances indicate better performance; a perfectly estimated configuration has zero distance to the truth. Boxplots of distances between the estimated changepoint configuration and the true configuration over the $100$ simulations are summarized in Figure \ref{fig:BS_CounterExample_Boxplots}. The red dots in the boxplots demarcate the average distance to the correct changepoint configuration. The boxplots show that binary segmentation underperforms all penalized likelihood methods.

\vspace{.12in} \noindent {\bf Best Practice 4: Use penalized likelihood MCPT methods --- or at least one of the binary segmentation improvements like wild binary segmentation or wild contrast maximization.}

In fitting a penalized likelihood MCPT model, the autocorrelation structure of the series is  estimated in the fit.  Binary segmentation does not give such estimates, but they are not difficult to obtain after the piecewise regime means are subtracted from the series.  Some MCPT techniques only allow special time series structures. For example, \cite{Chakar_2017_AR1} requires AR$(1)$ errors.  While the AR order is not believed to be as important as other issues in most climate applications, it is also infeasible that an AR(1) correlation structure works for all series. See \cite{Hewa_2018} for daily temperature homogenization via penalized likelihoods. While \cite{Cho-2020-WCM} allows general AR($p$) errors, simulations indicate that wild contrast maximization tends to estimate too many changepoints. In what follows, we concentrate on general penalized likelihood techniques estimated by a genetic algorithm.

\subsection{Atlanta Airport Temperatures}

\begin{figure}
	\centering
	\includegraphics[scale=0.6]{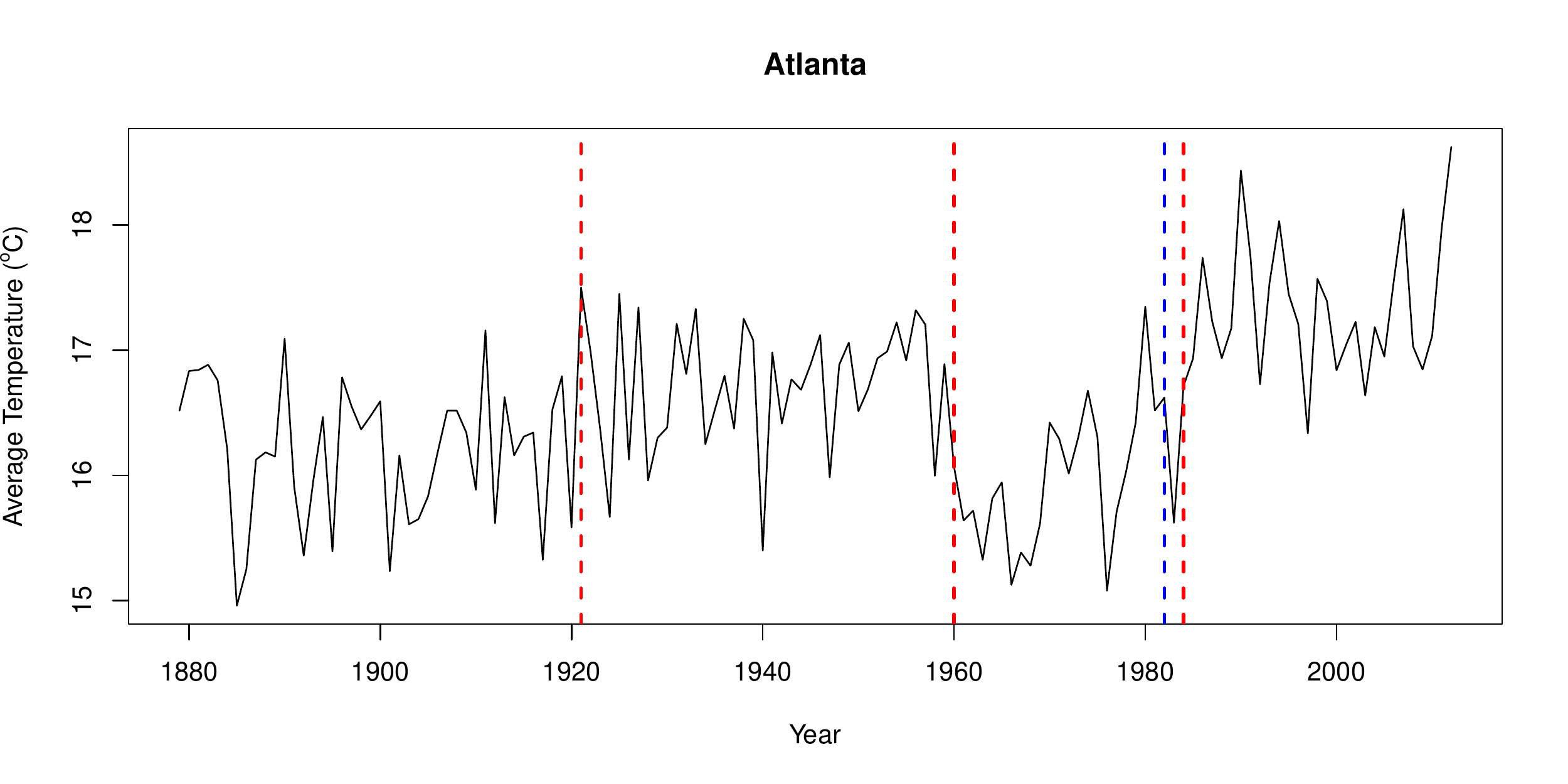}
	\caption{Changepoints flagged by a BIC penalized likelihood (red) and binary segmentation (blue). Binary segmentation flags one changepoint, while a BIC penalized likelihood flags three.}
	\label{fig:BS_CounterExample_Atlanta}
\end{figure}

To see differences between the approaches in practice, annual mean surface temperatures from 1879-2013 at Atlanta, Georgia's Hartsfield International Airport station will be analyzed. This dataset was provided by Berkeley Earth at \url{http://berkeleyearth.lbl.gov/station-list/}, and reflects ``raw'' temperatures that were not adjusted for potential artifacts. BIC penalized likelihoods and binary segmentation were fitted and compared, with results depicted in Figure \ref{fig:BS_CounterExample_Atlanta}. Binary segmentation flags only one changepoint in the early 1980s, while a BIC penalized likelihood approach estimates three changepoints, occurring in the 1920s, 1960s, and 1980s.  Our binary segmentation algorithm uses the SCUSUM AMOC test with a $95\%$ confidence threshold and accounts for autocorrelation via an AR(1) model.  While detailed simulations illustrating the inferiority of binary segmentation are supplied in \cite{Shi_etal_2021}, binary segmentation often has trouble identifying mean shifts that move the series in opposite directions. In this case, the successive changepoints estimated by penalized likelihood move the series up, down, and then up, two changepoints are apparently missed by binary segmentation.

\subsection{Ignoring Trends}

As in the AMOC case, ignoring trends in the MCPT setting will produce spurious results.   
For if the trend $\mu_t=E[X_t]$ is decisively increasing or decreasing, but ignored in the analysis, then MCPT procedure should flag one or more changepoints in an attempt to shift with the series mean.  In the AMOC case, each different form of $\{ \mu_t \}$ changes the asymptotic percentiles of the statistical test \citep{Tang_MacNeill_1993}. In the MCPT case, as long as the trend has the same form and parameters in all series subsegments, the penalties in (\ref{eqn:pform}) can be used. Should one want models where all parameters shift at the changepoint times --- an example would allow the trend slope to depend on the regime --- then the penalties in (\ref{eqn:pform}) must be modified. The reader is referred to \cite{Shi-2022-CET} for the appropriate penalties.

\vspace{.12in} \noindent {\bf Pitfall 5: Applying MCPT techniques to series with trends or seasonality that neglects these features.}

Similarly to the AMOC techniques in Pitfall 2, applying a MCPT technique that neglects trends and seasonality can result in spurious changepoint declarations. For example, an increasing trend will likely be estimated as a series of changepoints acting as a stairway up.

\vspace{.12in} \noindent {\bf Best Practice 5:  Allow for trends and/or seasonality in series with these features.}

\subsection{Arctic Sea Ice}

\begin{figure}
	\centering
	\includegraphics[scale=0.6]{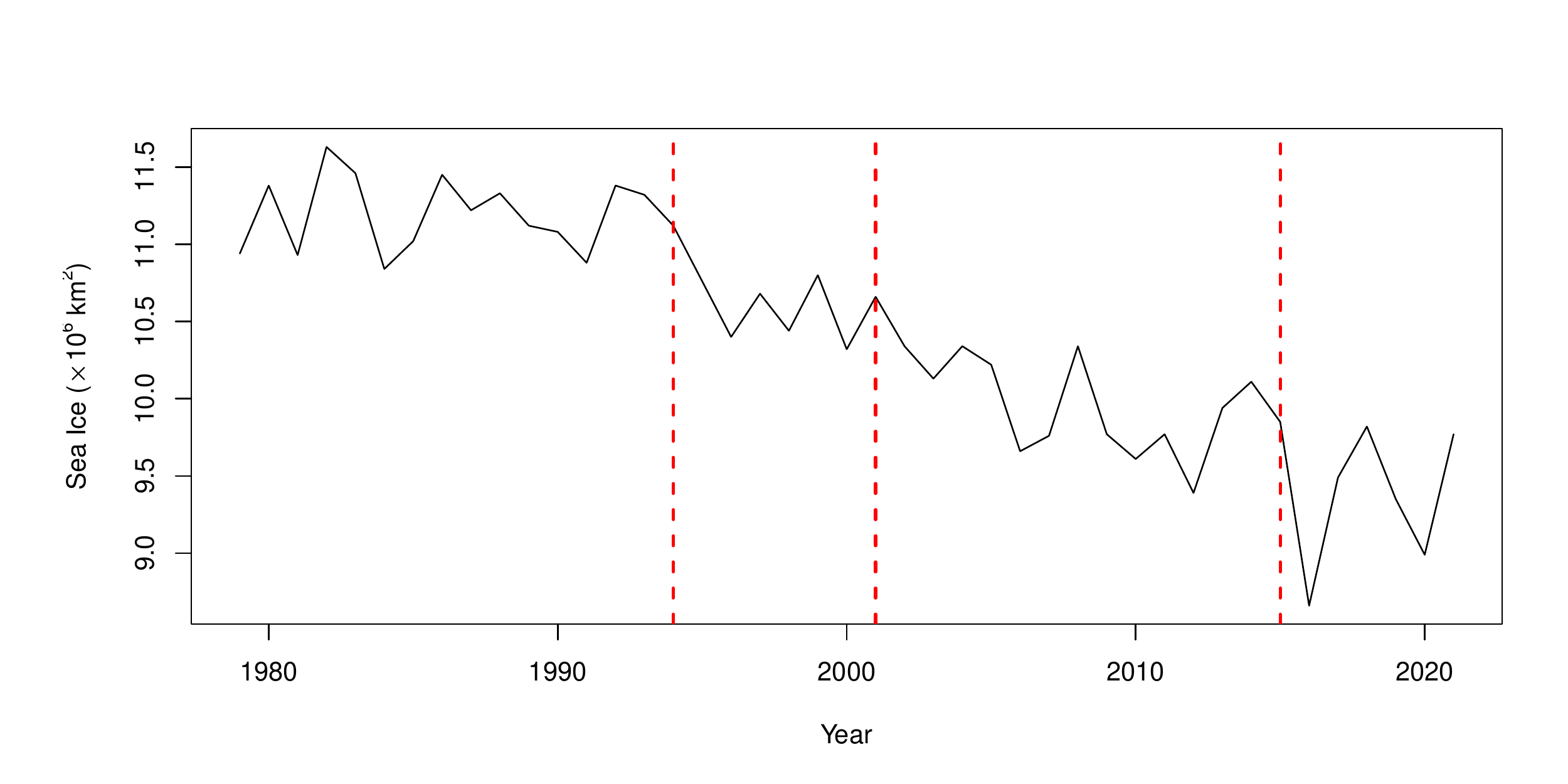}
	\caption{A changepoint analysis of the Arctic sea ice series.  When trends are ignored and AR$(1)$ errors are assumed, binary segmentation flags numerous mean shift changepoints in $1994,2001, 2015$ that attempt to ``follow the trend''. }
	\label{fig: Seaice-BinSeg}
\end{figure}

\begin{figure}
	\centering
	\includegraphics[scale=0.6]{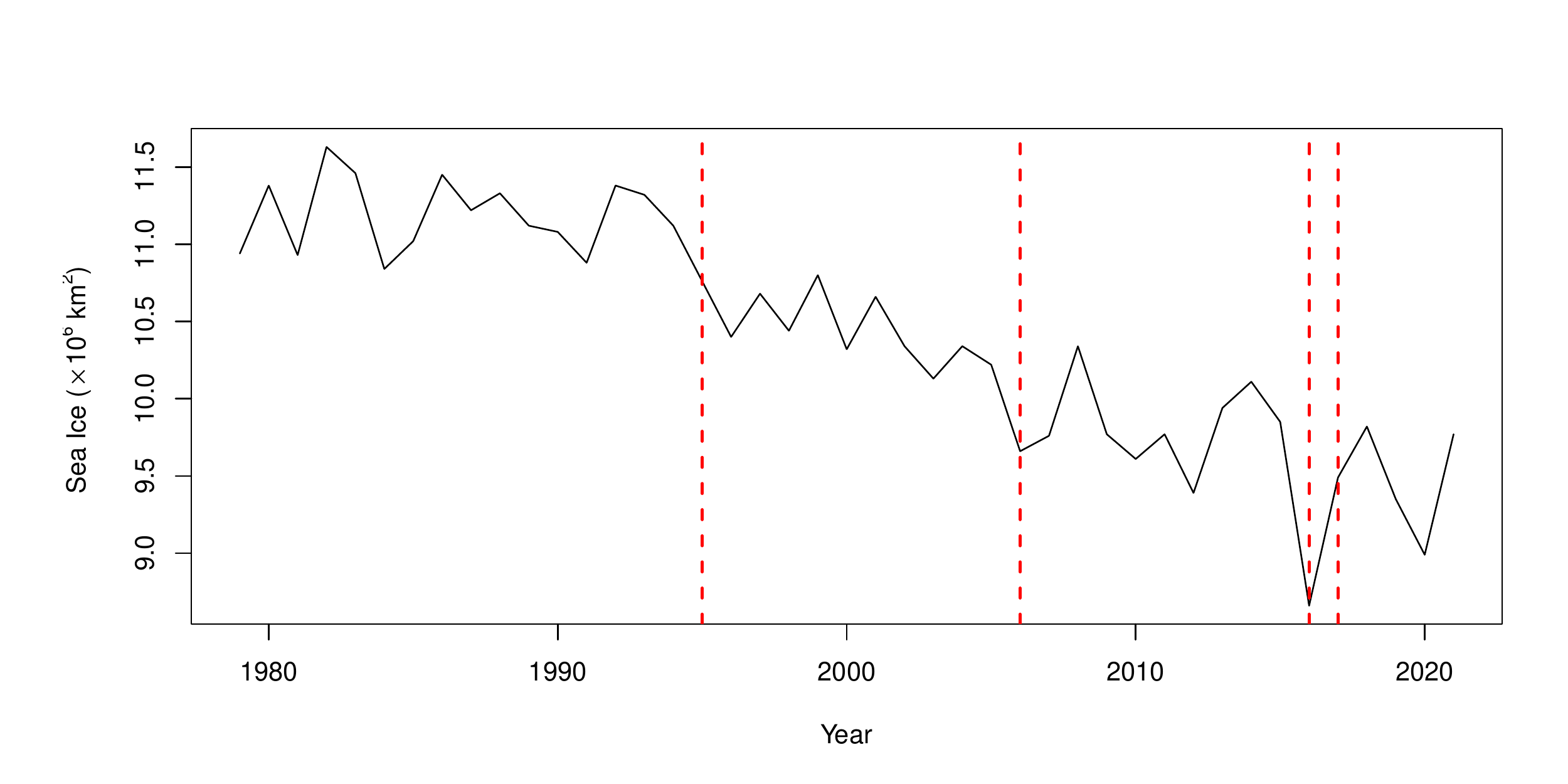}
	\caption{A changepoint analysis of the Arctic sea ice series.  When trends are ignored and AR$(1)$ errors are assumed, BIC penalized likelihood flags numerous mean shift changepoints in $1995, 2006, 2016, 2017$ that attempt to ``follow the trend''.  }
	\label{fig:Seaice-BIC-AR1}
\end{figure}

\begin{figure}
	\centering
	\includegraphics[scale=0.6]{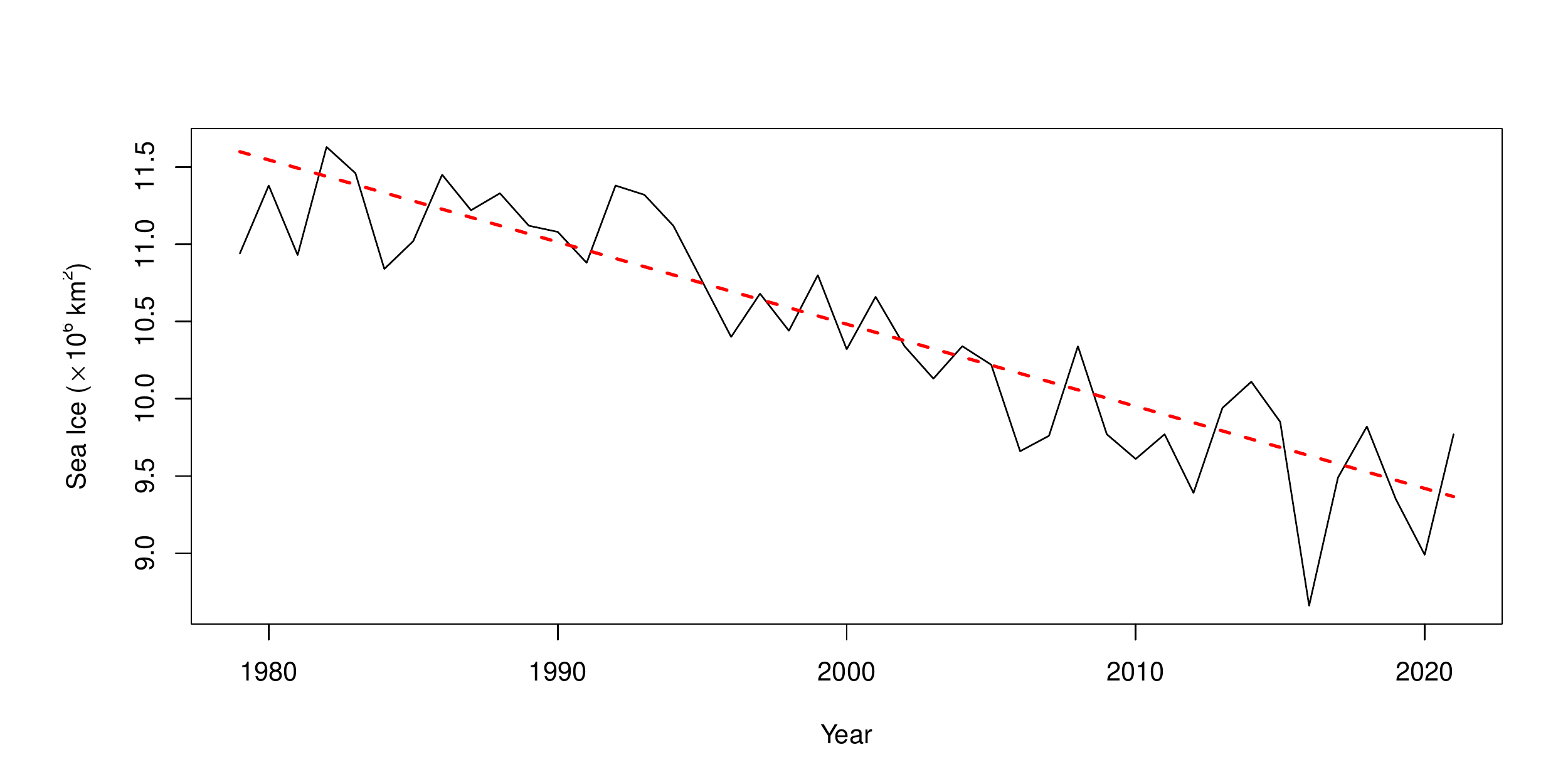}
	\caption{After a linear decreasing trend is added to the analysis and AR$(1)$ errors are still assumed, all changepoints are repealed in a BIC penalized likelihood fit.}
	\label{fig:Seaice-BIC-Trend+AR1}
\end{figure}

To illustrate the importance of accounting for trends, we analyse a series of September sea ice extent in the Northern hemisphere from 1979 - 2021. The data was provided by the National Snow and Ice Data Center, and downloaded from: \url{https://nsidc.org/data}. The series exhibits a decreasing trend that is partly attributable to increasing greenhouse gases emissions \citep{Meredith_2019}.  Figure \ref{fig: Seaice-BinSeg}, \ref{fig:Seaice-BIC-AR1} and \ref{fig:Seaice-BIC-Trend+AR1} show the series and some MCPT fits.  Figure \ref{fig: Seaice-BinSeg} is a binary segmentation fit with AR(1) errors that estimates three changepoints.  A BIC-type penalized likelihood estimated MCPT configuration with AR(1) errors identifies four changepoints (Figure \ref{fig:Seaice-BIC-AR1}).  Both of these fits were calculated assuming no trend.  When a linear trend is allowed in the BIC penalized likelihood, all changepoints are repealed (Figure \ref{fig:Seaice-BIC-Trend+AR1}).  The estimated trend slope of sea ice retreat was $-0.053$ million km$^2$/year.

\section{Discussion and Comments} 

This paper highlights some common pitfalls in changepoint analysis/homogenization methods and suggests best practices to avoid them.  In general, changepoint methods are not robust to changes in the mean structure of a series and care is needed in their proper application.  Issues considered in the paper include correlation, trends, distributions of maximum statistics, and the type of multiple changepoint analysis employed.  The general mantra is that if a series feature is not obvious (say existence of trends or correlation), it is best to put that feature in a model and let statistical methods discern whether that feature truly exists. While the paper attempts to put forth a best practice, any user of changepoint methods in the climate sciences should be aware of the litany of mistaken or dubious analyses in the field.   For example, the number of changepoint declarations that would be repealed due to failure to consider positive autocorrelation would be extensive.

It is worth rehashing target minus reference series analyses versus target series analyses only (absolute versus relative homogenization). While the statistical procedure to analyse both settings are the same, subtraction of a references series often reduces trends and/or seasonal cycles, making some issues clearer.  Nonetheless, as shown here, formation of a target minus reference series often does not reduce series autocorrelation, nor need it totally eliminate trends and/or seasonal cycles.   Existence of metadata is another issue.   While most authors tend to eschew metadata in their analyses, \cite{Li_Lund_2015} show how an informative Bayesian prior can be constructed from it and used with an MDL penalized likelihood to enhance changepoint detection power.

Multiple changepoint techniques are actively being researched in statistics.  Computational advances are expected within the next few years, especially in regard to penalized likelihood methods \citep{GD_PELT}.  Other aspects about the problem are also being studied.  One thing that is already clear from the literature is the inferiority of ordinary binary segmentation techniques in multiple changepoint problems.   Here, we urge researchers to use better methods.

\vspace{1cm}

\section*{Acknowledgement}
Robert Lund thanks National Science Foundation Grant DMS-2113592 for partial support. Xueheng Shi thanks National Science Foundation Grant CCF-1934568 for partial support. Claudie Beaulieu thanks National Science Foundation Grant AGS-2143550 for support.

%
%
\section*{Data statement}
The Central England data used in this study is available at \url{https://www.metoffice.gov.uk/hadobs/hadcet/}.  We used the annual records from 1659-2020.

\bibliographystyle{unsrt}


\end{document}